\documentclass[twocolumn]{aastex61}
\shorttitle{Discovery of MAXI J1807$+$132}
\shortauthors{Shidatsu et al.}

\begin{document}

\title{Discovery of the new X-ray transient MAXI J1807$+$132: a Candidate of a Neutron Star Low-mass X-ray binary}

\correspondingauthor{Megumi Shidatsu}
\email{megumi.shidatsu@riken.jp}

\author{Megumi Shidatsu}
\affiliation{MAXI team, RIKEN, 2-1 Hirosawa, Wako, Saitama 351-0198, JAPAN}
\author{Yutaro Tachibana}
\affiliation{Department of Physics, Tokyo Institute of Technology, 2-12-1 Ookayama, Meguro-ku, Tokyo 152-8551}
\author{Taketoshi Yoshii}
\affiliation{Department of Physics, Tokyo Institute of Technology, 2-12-1 Ookayama, Meguro-ku, Tokyo 152-8551}
\author{Hitoshi Negoro}
\affiliation{Department of Physics, Nihon University, 1-8-14 Kanda-Surugadai, Chiyoda-ku, Tokyo 101-8308, Japan}
\author{Taiki Kawamuro}
\affiliation{Department of Astronomy, Kyoto University, Kitashirakawa-Oiwake-cho, Sakyo-ku, Kyoto, Kyoto 606-8502, Japan}
\author{Wataru Iwakiri}
\affiliation{MAXI team, RIKEN, 2-1 Hirosawa, Wako, Saitama 351-0198, JAPAN}
\author{Satoshi Nakahira}
\affiliation{MAXI team, RIKEN, 2-1 Hirosawa, Wako, Saitama 351-0198, JAPAN}
\author{Kazuo Makishima}
\affiliation{MAXI team, RIKEN, 2-1 Hirosawa, Wako, Saitama 351-0198, JAPAN}
\author{Yoshihiro Ueda}
\affiliation{Department of Astronomy, Kyoto University, Kitashirakawa-Oiwake-cho, Sakyo-ku, Kyoto, Kyoto 606-8502, Japan}
\author{Nobuyuki Kawai}
\affiliation{Department of Physics, Tokyo Institute of Technology, 2-12-1 Ookayama, Meguro-ku, Tokyo 152-8551}
\author{Motoko Serino}
\affiliation{MAXI team, RIKEN, 2-1 Hirosawa, Wako, Saitama 351-0198, JAPAN}
\author{Jamie Kennea}
\affiliation{Department of Astronomy and Astrophysics, 0525 Davey Laboratory, Pennsylvania State University, University Park, PA 16802, USA}

\begin{abstract}
We report on the detection and follow-up multi-wavelength observations 
of the new X-ray transient MAXI J1807$+$132 with the MAXI/GSC, 
{\it Swift}, and ground-based optical telescopes. The source was first 
recognized with the MAXI/GSC on 2017 March 13. About a week later, 
it reached the maximum intensity ($\sim$10 mCrab in 2--10 keV), and 
then gradually faded in $\sim$10 days by more than one order of magnitude. 
Time-averaged {\it Swift}/XRT spectra in the decaying phase can be 
described by a blackbody with a relatively low temperature 
(0.1--0.5 keV), plus a hard power-law component with a photon 
index of $\sim$2. These spectral properties are similar to those 
of neutron star low-mass X-ray binaries (LMXBs) in their dim periods. 
The blackbody temperature and the radius of the emission region varied 
in a complex manner as the source became dimmer.
The source was detected in the optical wavelength on March 27--31 as well. 
The optical flux decreased monotonically as the X-ray flux decayed. The 
correlation between the X-ray and optical fluxes is found to be consistent with 
those of known neutron star LMXBs, supporting the idea that the source 
is likely to be a transient neutron star LMXB. 
\end{abstract}

\keywords{X-rays: individual (MAXI J1807$+$132) --- X-rays: binaries}

\section{Introduction} \label{sec:intro}
Many of X-ray point sources in the sky 
have significant variability on various 
timescales. In particular, transient 
low-mass X-ray binaries (LMXBs), involving 
accreting neutron stars or black holes, 
exhibit dramatic outbursts, 
changing their X-ray luminosity by orders 
of magnitude in relatively short periods 
\citep[a few days to months; see e.g.,][for a review]{don07}.
Because their luminosity is mainly determined 
by the mass accretion rate, 
their transient nature makes them ideal 
objects to study the physics of accretion 
onto compact objects over a wide range of 
mass accretion rates. 

The new transient MAXI J1807$+$132, located 
at an off-Galactic-plane region with a 
Galactic lattitude of $15^\circ.5$, was first 
noticed on 2017 March 13 \citep{neg17} by the 
nova-search system \citep{neg16} of the Monitor 
of All-sky X-ray Image \citep[MAXI;][]{mat09}, 
in an X-ray image provided by the MAXI/Gas Slit 
Camera \citep[GSC;][]{mih11}. 
The source position estimated with the MAXI/GSC turned 
out to be consistent with that of 2MAXIt J1807$+$132, 
which is listed in the MAXI/GSC transient 
source catalog \citep{kaw16}, 
based on an X-ray flaring event detected 
in 2011 May.

\begin{figure}[htb]
\epsscale{1.1}
\plotone{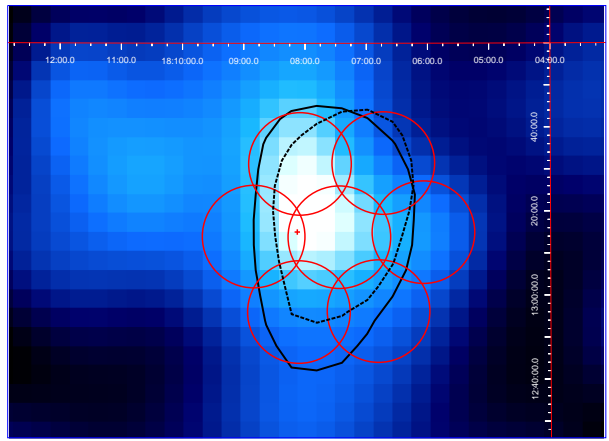}
\caption{A 2--4 keV image ($2^\circ.3$$\times$$1^\circ.7$) 
obtained in the MAXI/GSC observation from 2017 March 13 
to April 1. The thick solid and dashed lines (black) indicate 
the error regions of MAXI J1807$+$132 determined by the 
MAXI/GSC in the 2--4 keV and 4--10 keV bands, respectively. 
The seven red circles represent the fields-of-view 
of the {\it Swift}/XRT for individual pointings 
on March 26. The red cross point indicates the 
position of the source detected with {\it Swift}.
\label{fig:Ximage}}
\end{figure}

As shown in Figure~\ref{fig:Ximage}, the 
source was detected in a 7-tile follow-up 
observation with the {\it Swift}/X-ray 
Telescope \citep[XRT;][]{geh05}, and 
UltraViolet and Optical Telescope 
\citep[UVOT;][]{rom05}, in the X-ray band 
and the optical to ultraviolet bands, respectively. 
Thus, the source position was determined 
accurately as $(\alpha^{2000},$ $\delta^{2000}) = 
(18^{\mathrm h}08^{\mathrm m}07^{\mathrm s}.549, 
+13^\circ15'05.''40)$ and ($l$, $b$) 
$= (40^\circ.123127,$ $15^\circ.501653)$ 
with a 90 \% uncertainty of $0.''16$ 
\citep{ken17a, ken17b}. The UVOT $u$, $b$, and $v$-band 
magnitudes were $17.4 \pm 0.1$ mag, $18.4 \pm 0.2$ 
mag, and $>$17.6 mag at that time \citep{ken17b}, 
respectively. The source was also followed by ground-based 
optical telescopes \citep{shie17,tac17,mun17,arm17a,kon17}.

These follow-up X-ray and optical observations 
have provided pieces of information to 
understand the nature of the object. 
The source showed significant flux variations 
in the optical band \citep{shie17,tac17,arm17a,kon17} 
as well as X-rays \citep{neg17,arm17a,kon17}. 
\citet{den17} recognized the optical counterpart 
in an image taken about 35 years ago and 
suggested past activities of the source. The source 
was also detected optically in PanSTARRS-1 
multi-epoch data, with a magnitude which was 
larger by 2--3 (i.e., $\sim$10 times fainter 
in terms of the flux) than those obtained 35 years 
ago and in 2017 March \citep{den17}. 
LMXB-like characteristics were found through 
preliminary modeling of the XRT spectra 
\citep{shi17} and optical spectroscopy 
\citep{mun17}.

In this paper, we present the first results from 
the multi-wavelength monitoring of 
MAXI J1807$+$132 in 2017 March and April,
with the MAXI/GSC, {\it Swift} and ground-based 
optical telescopes, and discuss the nature of the 
source. Throughout the paper, errors represent 
the 90\% confidence intervals of a single parameter 
with $\Delta \chi^2 = 2.706$, unless otherwise 
stated.

\section{X-ray Observations and Results} \label{sec:xobs}

\begin{deluxetable}{ccccc}[tbh]
\tablecaption{Log of {\it Swift}/XRT observations\label{tab:xobslog}}
\tablecolumns{5}
\tablenum{1}
\tablewidth{0pt}
\tablehead{
\colhead{Date} &
\colhead{Start time} &
\colhead{End time} &
\colhead{Net exposure} & 
\colhead{XRT} \\
\colhead{} &
\colhead{(UT)} &
\colhead{(UT)} &
\colhead{(ks)} & 
\colhead{mode\tablenotemark{a}}
}
\startdata
2017 Mar 26 & 08:44:23 & 10:26:52 & 0.22 & PC \\
2017 Mar 27 & 07:01:39 & 09:50:56 & 1.94 & WT \\
2017 Mar 29 & 05:08:51 & 07:01:38 & 1.99 & WT \\
2017 Mar 31 & 08:13:39 & 11:12:56 & 2.05 & WT \\
2017 Apr 02 & 14:18:11 & 14:29:56 & 0.69 & WT \\
2017 Apr 04 & 11:06:57 & 11:07:02 & 0.005 & WT \\
2017 Apr 05 & 15:36:52 & 15:52:53 & 0.95 & PC \\
2017 Apr 06 & 06:08:03 & 06:16:52 & 0.51 & PC \\
2017 Apr 07 & 21:38:20 & 21:54:53 & 0.98 & PC \\
2017 Apr 08 & 21:29:20 & 21:44:53 & 0.91 & PC \\
2017 Apr 09 & 16:56:20 & 17:12:53 & 0.99 & PC \\
\enddata
\tablenotetext{a}{PC and WT indicate the Photon 
Counting mode and the Windowed Timing mode, respectively.}
\end{deluxetable}

\begin{figure}[htb]
\epsscale{1.1}
\plotone{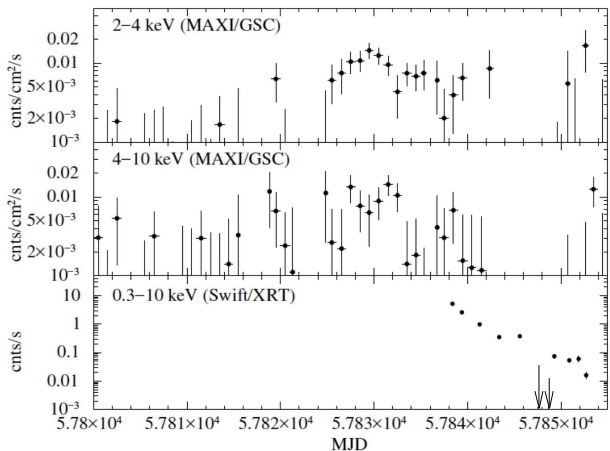}
\caption{MAXI/GSC light curves in 
2--4 keV (top) and 4--10 keV (middle) with one-day 
time bins, and {\it Swift/XRT} light curve in 0.3--10 keV (bottom), 
with one observation per bin. The arrows represent upper limits. 
MJD 57825 corresponds to 2017 March 13.
\label{fig:MAXIlc}}
\end{figure}

\subsection{Long-term Monitoring with MAXI}\label{lc}
After the first detection on 2017 March 13, 
MAXI J1807$+$132 has been monitored with 
the MAXI/GSC. Figure~\ref{fig:MAXIlc} 
presents background-subtracted 2--4 keV and 4--10 keV 
light curves, created through the method of 
image fitting \citep{mor16} to the MAXI/GSC event 
data version 1.8. The source intensity 
reached a maximum at around MJD 57827--MJD 57832 
(2017 March 15--20), and then gradually decreased.
The averaged 4--10 keV intensity was $\approx$9 
mCrab in that period, which is 
comparable with the peak intensity of the X-ray 
flaring event from 2MAXIt J1807$+$132 recorded 
in \citet{kaw16}.

\begin{deluxetable*}{lcccccccc}[tbh]
\tablecaption{Best-fit parameters of various spectral 
models for the {\it Swift}/XRT data on March 27 and 
31 \label{tab:Xfit}}
\tablecolumns{9}
\tablenum{1}
\tablewidth{0pt}
\tablehead{
\colhead{Parameter} 
& \colhead{$N_\mathrm{H}$}
& \colhead{$\Gamma$ or $\tau$} 
& \colhead{$N_\mathrm{pl}$\tablenotemark{a}}
& \colhead{$R_\mathrm{bb~(comp)}$\tablenotemark{b}}
& \colhead{$kT_\mathrm{bb}$ or $kT_\mathrm{in}$} 
& \colhead{$R_\mathrm{bb}$ or $R_\mathrm{in}$\tablenotemark{c}}
& \colhead{$F_\mathrm{X}$\tablenotemark{d}} 
& \colhead{$\chi^2/\mathrm{dof}$\tablenotemark{e}} \\
\colhead{Unit} 
& \colhead{$10^{21}$ cm$^{-2}$}
& \colhead{} 
& \colhead{}
& \colhead{km}
& \colhead{keV}
& \colhead{km}
& \colhead{erg s$^{-1}$ cm$^{-2}$}
& \colhead{}
}
\startdata
\multicolumn{9}{l}{Model: {\tt TBabs*powerlaw}} \\
\tableline 
Mar. 27 & $2.5 \pm 0.6$ & $2.4^{+0.2}_{-0.1}$ & $12 \pm 1$ 
& - & - & - & $5.7 \times 10^{-11}$ & $63/65$ \\
Mar. 27 & $1.0$~(fixed) & $2.07 \pm 0.07$ & $8.4 \pm 0.4$ 
& - & - & - & $4.6 \times 10^{-11}$ & $84/66$ \\
Mar. 31 & $<1.3$ & $3.2^{+1.0}_{-0.4}$ & $0.7^{+0.4}_{-0.1}$  & - & - & - & $3.9 \times 10^{-12}$ & $19/18$ \\
Mar. 31 & $1.0$~(fixed) & $3.9 \pm 0.5$ & $1.0 \pm 0.2$ 
& - & - & - & $8.2 \times 10^{-12}$ & $21/19$ \\
\tableline
\multicolumn{9}{l}{Model: {\tt TBabs*(diskbb+powerlaw)}} \\
\tableline 
Mar. 27 & $1.0$~(fixed) & $1.6^{+0.3}_{-0.7}$ & $3 \pm 2$ & - & $0.50^{+0.09}_{-0.06}$ & $1.8^{+0.3}_{-0.7}$ & $4.1 \times 10^{-11}$ & $60/64$ \\
Mar. 31 & $1.0$~(fixed) & $1.4^{+1.0}_{-1.3}$ & $0.2^{+0.3}_{-0.1}$ & - & $0.15 \pm 0.03$ 
& $19^{+17}_{-9}$ & $7.8 \times 10^{-12}$ & $10/17$ \\
\tableline 
\multicolumn{9}{l}{Model: {\tt TBabs*(bbodyrad+powerlaw)}} \\
\tableline 
Mar. 27 & $1.0$~(fixed) & $1.8 \pm 0.2$ & $5 \pm 1$  & - & $0.32^{+0.06}_{-0.04}$ & $4.2^{+1.6}_{-1.5}$ & $4.1 \times 10^{-11}$ & $59/64$ \\
Mar. 31 & $1.0$~(fixed) & $1.5 \pm 1.0$ & $0.2^{+0.3}_{-0.1}$ & - & $0.12 \pm 0.02$ 
& $32^{+22}_{-12}$ & $7.2 \times 10^{-12}$ & $10/17$ \\
\tableline 
\multicolumn{9}{l}{Model: {\tt TBabs*(bbodyrad+compps(bbodyrad))}} \\
\tableline 
Mar. 27\tablenotemark{f} & $1.0$~(fixed) & $2.5^{+0.5}_{-0.4}$~\tablenotemark{h} & - & $18^{+1}_{-3}$ & $0.22 \pm 0.05$ & $<10$ & $3.7 \times 10^{-11}$ & $62/64$ \\
Mar. 27\tablenotemark{g} & $1.0$~(fixed) & $0.59^{+0.08}_{-0.07}$ & - & $12.9^{+0.2}_{-1.7}$ & $0.26 \pm 0.03$ & $<6$ & $3.8 \times 10^{-11}$ & $64/64$ \\
Mar. 31\tablenotemark{f} & $1.0$~(fixed) & $3.0^{+0.0}_{-1.2}$~\tablenotemark{h} & - & $7 \pm 0.2$ & $0.11 \pm 0.02$ & $33^{+21}_{-12}$ & $6.3 \times 10^{-12}$ & $10/17$ \\
Mar. 31\tablenotemark{g} & $1.0$~(fixed) & $1.5^{+1.5}_{-1.3}$~\tablenotemark{h} & - & $7 \pm 0.2$ & $0.12 \pm 0.02$ & $30^{+20}_{-9}$ & $6.4 \times 10^{-12}$ & $10/17$ \\
\tableline 
\multicolumn{9}{l}{Model: {\tt TBabs*(diskbb+compps(diskbb))}} \\
\tableline 
Mar. 27\tablenotemark{f} & $1.0$~(fixed)& $3.0^{+0.0}_{-0.8}$~\tablenotemark{h} & - & $1.1^{+0.3}_{-0.1}$ & $0.41^{+0.07}_{-0.12}$ & $<2.9$ & $3.9 \times 10^{-11}$ & $60/64$ \\
Mar. 27\tablenotemark{g} & $1.0$~(fixed) & $0.7^{+2.3}_{-0.3}$~\tablenotemark{h} & - & $1.2 \pm 0.2$ & $0.44^{+0.10}_{-0.08}$ & $<2.5$ & $4.0 \times 10^{-11}$ & $59/64$ \\
Mar. 31\tablenotemark{f} & $1.0$~(fixed) & $3.0^{+0.0}_{-1.1}$~\tablenotemark{h} & - & $2.0^{+0.7}_{-0.2}$ & $0.14 \pm 0.03$ & $20^{+12}_{-9}$ & $6.8 \times 10^{-12}$ & $11/17$ \\
Mar. 31\tablenotemark{g} & $1.0$~(fixed) & $2.0^{+1.0}_{-1.7}$~\tablenotemark{h} & - & $1.8^{+0.7}_{-0.4}$ & $0.15 \pm 0.03$ & $17^{+16}_{-8}$ & $6.9 \times 10^{-12}$ & $10/17$ \\
\enddata
\tablenotetext{a}{Normalization of the power-law component, 
in units of $10^{-3}$ photons keV cm$^{-2}$}
\tablenotetext{b}{Radius of the emission region of the seed photons for Compton scattering, 
 calculated from the photon flux of the {\tt compps} component in 0.8--100 keV, by assuming 
 a distance of 5 kpc and a spherical corona.}
\tablenotetext{c}{The radius of emission region for the {\tt bbodyrad} component, or 
the inner disk radius for the {\tt diskbb} component. The distance and the inclination 
angle are assumed as 5~kpc and 0$^\circ$, respectively.}
\tablenotetext{d}{The unabsorbed X-ray flux in the 0.3--10 keV band.}
\tablenotetext{e}{C-statistic/dof is instead presented for the March 31 results.}
\tablenotetext{f}{$kT_\mathrm{e}= 20$ keV is assumed.}
\tablenotetext{g}{$kT_\mathrm{e}= 100$ keV is assumed.}
\tablenotetext{h}{The upper limit is pegged.}
\end{deluxetable*}

\subsection{{\it Swift}/XRT Spectral Analysis}\label{xrtana}
A series of follow-up pointed observations of MAXI J1807$+$132 
were carried out with {\it Swift} in the decaying phase. 
The net exposure was 0.5--2 ks in each observation. 
A log of these observations 
is given in Table~\ref{tab:xobslog}, and the 0.3--10 keV 
XRT light curve is shown in Figure~\ref{fig:MAXIlc}. 
We analyzed the XRT spectra taken during two weeks 
from the first {\it Swift} observation (March 26),  
using XSPEC version 12.9.1m \citep{arn96}. 
The light curve, spectra, and responses were downloaded 
via the online tools provided by the UK Swift Science Data 
Centre \citep{eva09}\footnote{\url{http://www.swift.ac.uk/user_objects/}}. 
The spectra taken on March 27 and March 29, 
which have the best statistics among the present 
XRT datasets, were binned so that at least 30 
counts are contained in each bin, and were analyzed 
on the basis of the $\chi^2$ statistics. The other 
data, which have lower statistics, were grouped so 
that each bin has at least one count, and the 
Cash-statistics \citep{cas79} were adopted in the 
spectral analysis. The data taken on April 4 and 5 
were omitted, because the source was not detected 
significantly. The data on April 6 and 7 were merged 
together to improve statistics; so were those on April 8 and 9. 
In the following analysis, the {\tt TBabs} model is 
employed as the interstellar absorption model, with 
the table of solar abundances provided by \citet{wil00}.

\begin{figure}[htb]
\epsscale{1.1}
\plotone{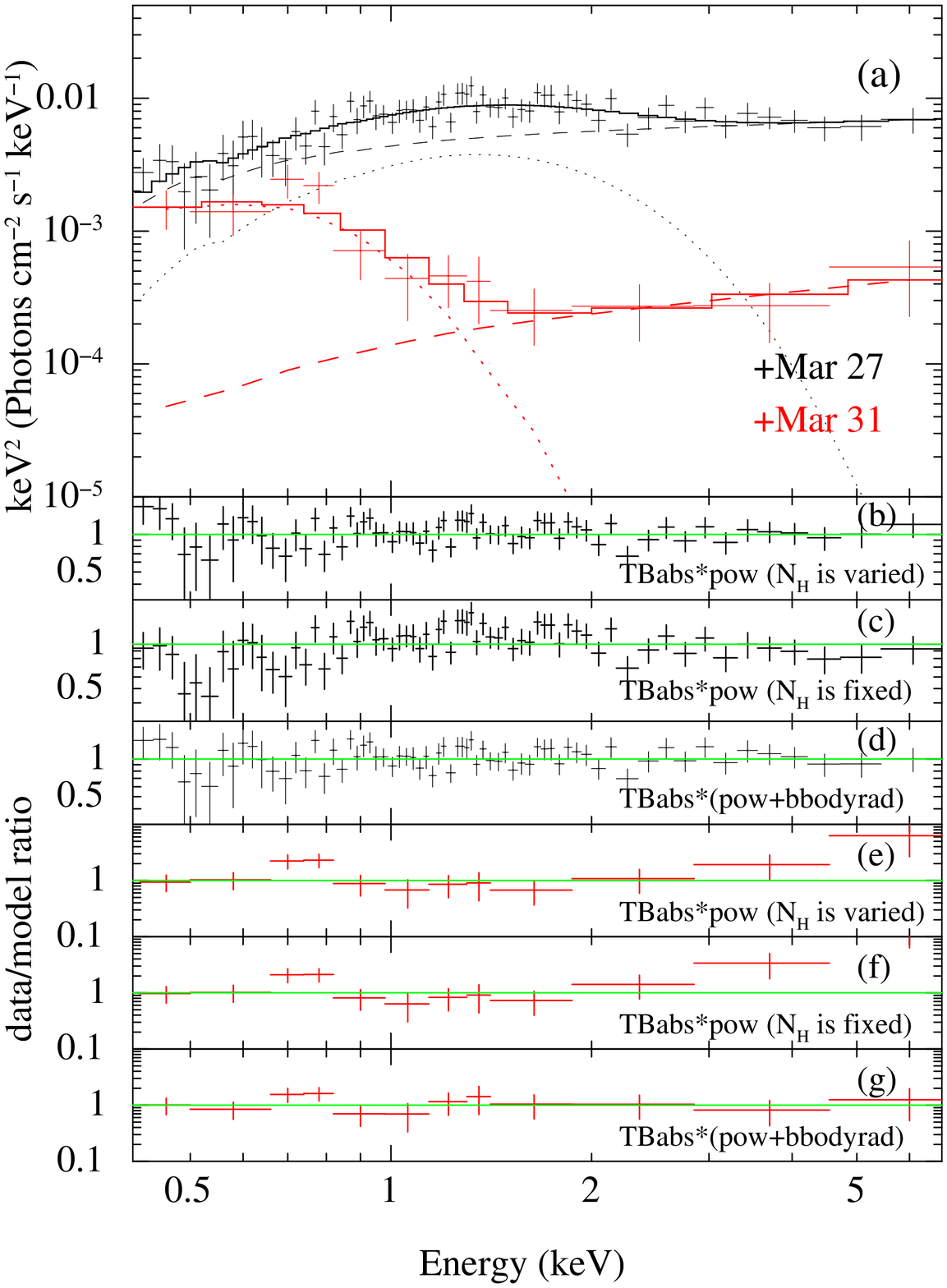}
\caption{(a) Time-averaged {\it Swift}/XRT spectra 
on March 27 (black) and 31 (red) in the $\nu F_\nu$ form, 
with their best-fit {\tt TBabs*(powerlaw+bbodyrad)} models. 
The contributions of the {\tt powerlaw} and {\tt bbodyrad} 
components are also shown separately in the 
dashed and dotted lines, respectively.
(b) The March 27 data divided by the 
{\tt TBabs*(powerlaw)} model with free $N_\mathrm{H}$, 
(c) by that with $N_\mathrm{H} = 1 \times 10^{21}$ cm $^{-2}$ (fixed), 
and (d) divided by the {\tt TBabs*(powerlaw+bbodyrad)} model.
(e)--(g) The same as (b)--(d), respectively, for the March 31 data.
The spectra and the ratio on March 31 
are binned for presentation purposes. 
\label{fig:XRTfit}}
\end{figure}

As shown in Figure~\ref{fig:XRTfit}, we first analyzed 
the XRT spectra on March 27 and 31, 
which were obtained when the source intensity 
was relatively high and low, respectively. 
The spectrum on March 27 can 
be fit with an absorbed power-law model with a 
photon index of $\Gamma=2.4^{+0.2}_{-0.1}$ and 
a column density of $N_{\mathrm H} = (2.5 \pm 0.6) 
\times 10^{21}$ cm$^{-2}$, in 
agreement with the first report by \citet{ken17a,ken17b}. 
This model is fully acceptable (Fig.~\ref{fig:XRTfit}b), 
with $\chi^2/\mathrm{dof}=63/65$. The resultant 
parameters are listed in Table~\ref{tab:Xfit}.
However, the estimated column density is somewhat 
higher than the total Galactic column, 
$N_{\mathrm H}= 1.0 \times 10^{21}$ 
cm$^{-2}$ derived using the tool {\tt nh} in HEASOFT version 
6.19, and that calculated from the optical spectrum 
of the source \citep[$N_{\mathrm H} \sim 1.6 
\times 10^{21}$ cm$^{-2}$;][]{mun17}. 
The fit quality became worse ($\chi^2/\mathrm{dof}=84/66$), 
when $N_{\mathrm H}$ is fixed at $1.0 \times 10^{21}$ 
cm$^{-2}$ (Fig.~\ref{fig:XRTfit}c).

As observed in Figure~\ref{fig:XRTfit}(a), the XRT spectrum 
changed significantly in shape from March 27 to March 31.
The March 31 spectrum appears to consist of a softer 
component dominant in $<2$ keV, and a harder 
power-law-like component with a photon index 
of $\leq$ 2 extending above 2 keV. Then, the spectral change 
from March 27 to 31 can be understood if the latter component 
decreased by an order of magnitude, with a relatively small change 
in the former. As a confirmation, we forced a single power-law 
model to the March 31 spectrum. The fit was formally acceptable 
(Table~\ref{tab:Xfit}), but as shown in Figure~\ref{fig:XRTfit}(e), 
the data in $>$ 2 keV systematically exceeded the best-fit 
power-law model, which was required to have a very steep 
($\Gamma \sim$ 3.2) slope. We hence regard the single 
power-law model as inappropriate, both on March 27 and March 
31.

Based on the above consideration, 
we next fitted the spectra with a model 
composed of a power-law 
component and a thermal emission component: a 
blackbody ({\tt bbodyrad} in XSPEC terminology) 
or a multi-color disk blackbody \citep[{\tt diskbb}][]{mit84}.
Here $N_{\mathrm H}$ is assumed as 
$1 \times 10^{21}$ cm$^{-2}$. 
Then, the spectrum on March 27 has been well 
described by the model, 
with the reduced chi-squared values of 
$\chi^2/\mathrm{dof} = 60/64$ and $\chi^2/\mathrm{dof}= 59/64$ 
in the case of {\tt bbodyrad} and {\tt diskbb}, 
respectively. These values are almost the same 
as that obtained with a single power-law model in
which $N_\mathrm{H}$ is left as a free parameter, 
while significantly smaller than that of the 
same model with $N_\mathrm{H} = 1 \times 10^{21}$ 
cm$^{-2}$. The XSPEC script {\tt simftest} shows 
an null-hypothesis probability of $<10^{-6}$ 
in the latter case. Both models gave relatively 
small photon indices: $\Gamma =1.8 \pm 0.2$ and 
$1.6^{+0.3}_{-0.7}$, with the {\tt bbodyrad} and 
{\tt diskbb} models, respectively. 

The combination of the thermal and power-law 
components has successfully described the 
spectrum on March 31 as well. 
The reduced chi-squared values were slightly 
reduced from that of the single power-law model 
with $N_\mathrm{H} = 1 \times 10^{21}$ cm$^{-2}$ 
($\Delta \mathrm{C}$-statistic $\approx 11$ for 
$\Delta \nu = 2$), 
and the residual structure above $\approx$2 keV 
dissapeared (Fig.~\ref{fig:XRTfit}g).  
The probability that the improvement is 
only due to a random fluctuation is 0.004, 
according to {\tt simftest}.
The temperature of the soft thermal component 
became lower by a factor of $\approx$3, 
and the apparent linear size of its emission region 
became larger by a factor of 8--10, than those on 
March 27 (see Table~\ref{tab:Xfit}).

\begin{figure*}[htb]
\epsscale{1.1}
\plottwo{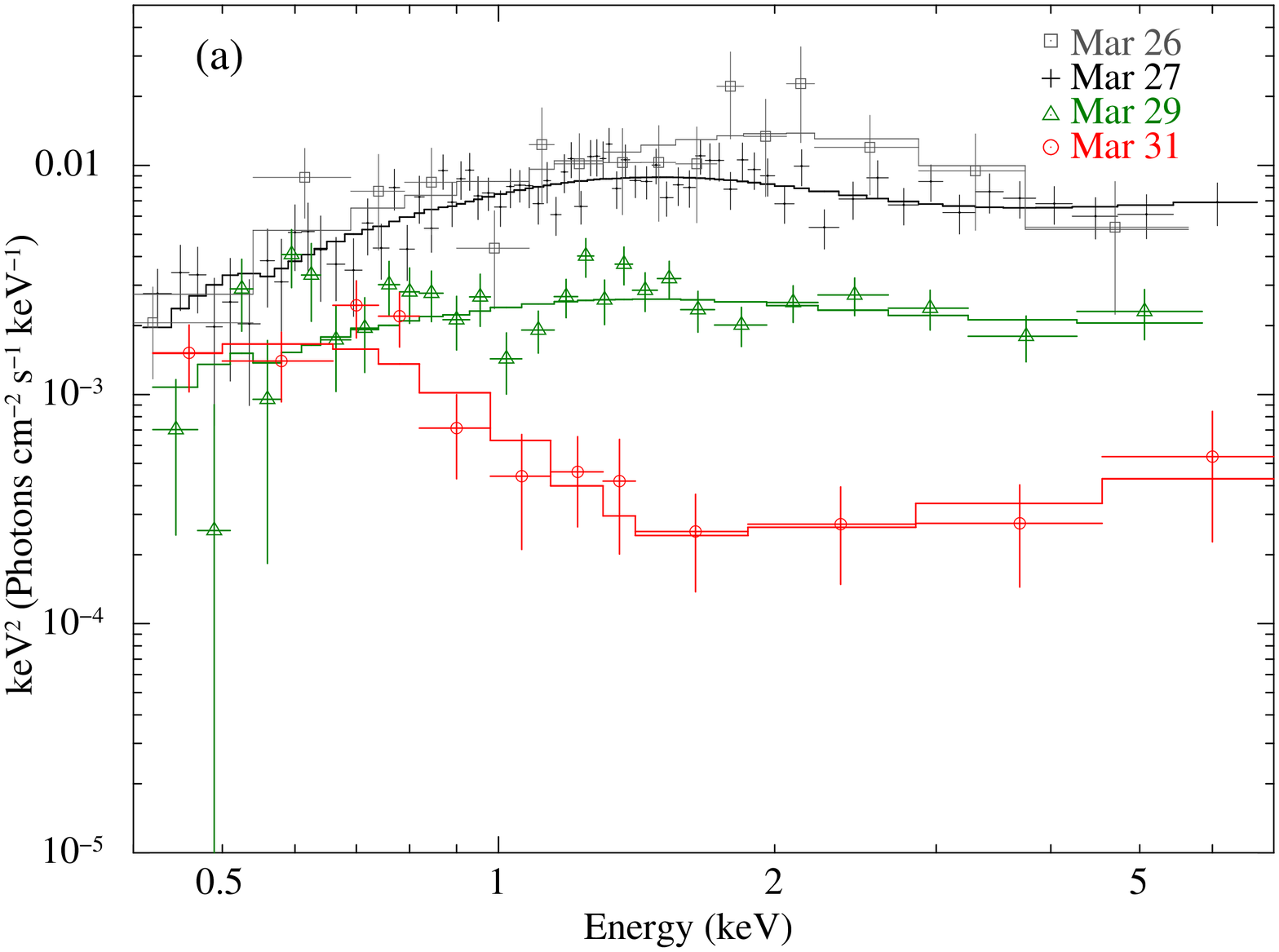}{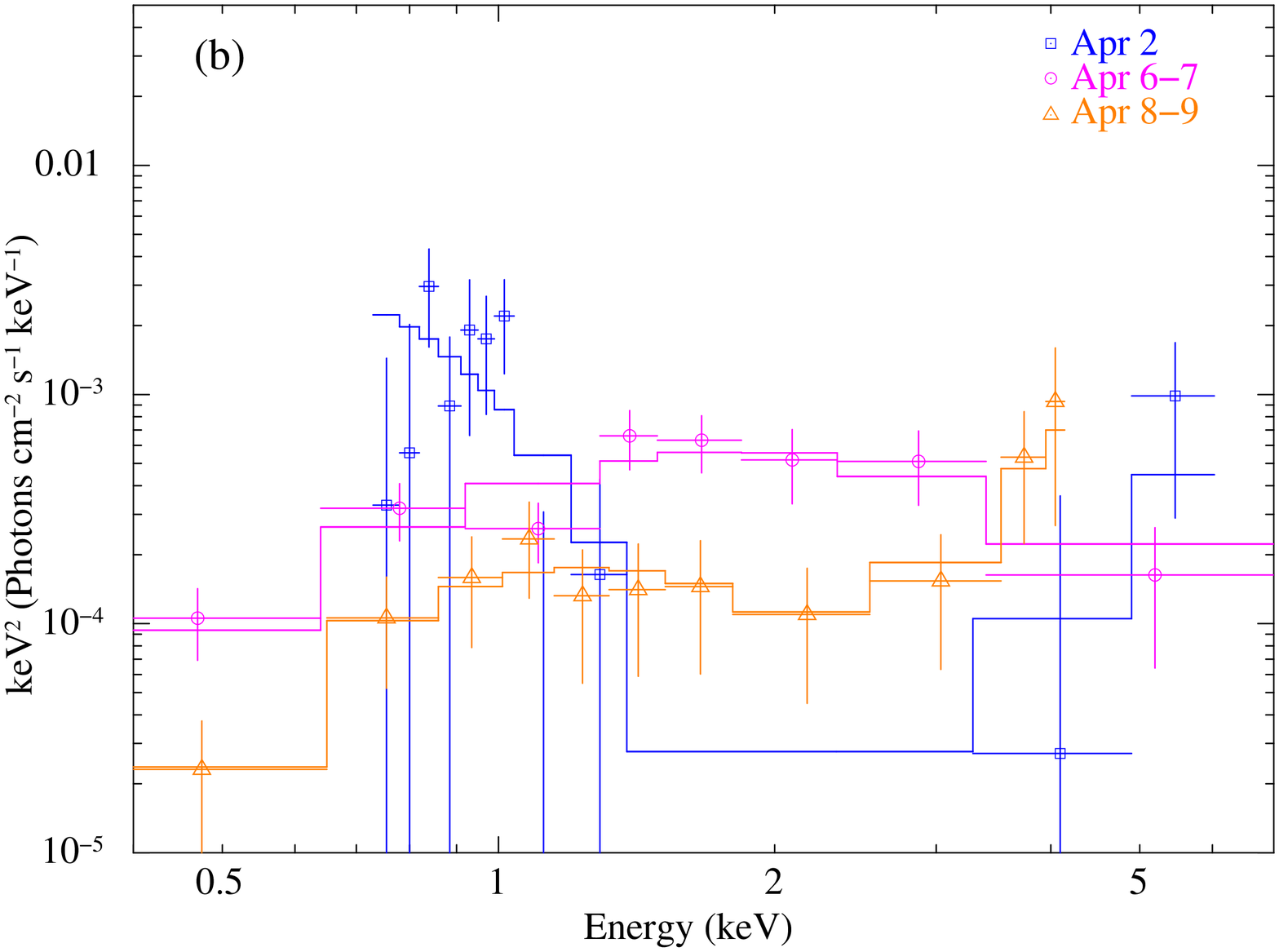}
\caption{Time-averaged {\it Swift}/XRT spectra in the 
individual epochs on (a) March and (b) April, in 
the $\nu F_\nu$ form, with their best-fit 
{\tt TBabs*(powerlaw+bbodyrad)} models. The March 
27 and 31 spectra in (a) are identical to those in 
Fig.~\ref{fig:XRTfit}.
\label{fig:XRTspecall}}
\end{figure*}

\begin{figure}[htb]
\epsscale{1.1}
\plotone{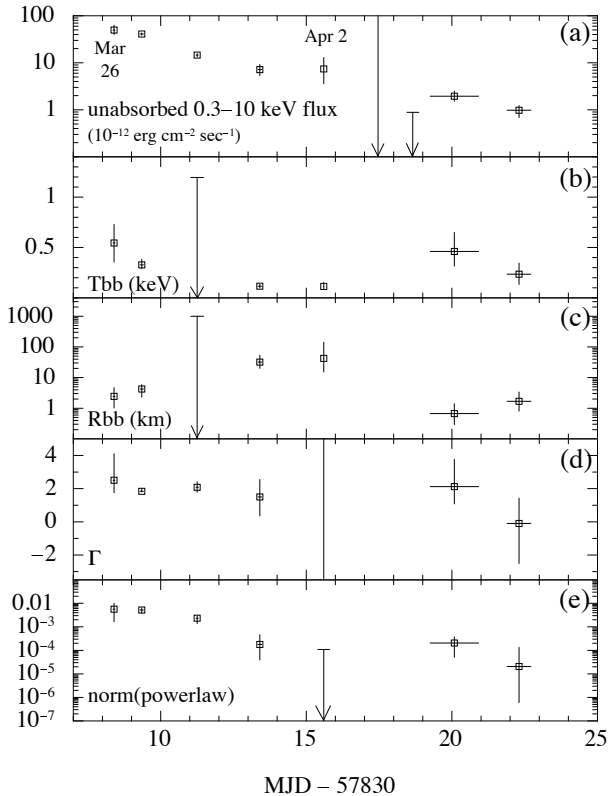}
\caption{Time evolution of the parameters of  
the {\tt TBabs*(bbodyrad+powerlaw)} model. (a) The 
unabsorbed 0.3--10 keV flux, (b) the temperature 
of the blackbody component, (c) the emission radius of 
the blackbody component, (d) the photon index, and 
(e) the normalization of the power-law component. 
The photon index on MJD 57845 (April 2) was 
not constrained. Only a loose flux limit ($< 1.3 \times 
10^{-8}$ erg cm$^{-2}$ s$^{-1}$) was obtained 
for MJD 57847 (April 4).\label{fig:trend_fitpars}}
\end{figure}

The soft X-ray component is most likely optically-thick 
thermal emission from the surface 
of a compact object (if it is not a black hole) 
or an accretion disk. The hard component, like in 
the present case, is often considered to originate via 
Comptonization of these thermal photons by a cloud 
of hot electrons \citep[e.g.,][]{don07,lin07}. 
Assuming that the observed hard tail is produced by 
Comptonization, we replaced the phenomenological 
power-law model with the {\tt compps} model \citep{pou96}.

The {\tt compps} model calculates a Comptonized 
spectrum when we specify the electron temperature 
$kT_\mathrm{e}$, the optical depth for scattering $\tau$, the energy 
distribution of the seed photons, and the geometry of the 
Comptonization cloud (or corona). In the present study, 
following previous works \citep[e.g.,][]{sak14}, 
a spherical corona ({\tt geom} $=4$) was assumed (but 
see below for the cases of different geometries), with only 
thermal electrons. 
We first tested the case where the seed photons are 
provided by blackbody emission. The model is expressed  
as {\tt TBabs*(compps+bbodyrad)}, where the seed photon 
temperature $kT_\mathrm{seed}$ of the {\tt compps} 
component was linked to $kT_\mathrm{bb}$ of the 
{\tt bbodyrad} component. 
Since the data did not allow us to simultaneously 
constrain $\tau$ and $kT_\mathrm{e}$, 
we fixed $kT_\mathrm{e}$ at 20 keV and 100 keV, 
and left  $\tau$ as a free parameter. The other free parameters 
in this model are $kT_\mathrm{bb}$ and the normalizations 
of the {\tt bbodyrad} and {\tt compps} components. 
We ignored the reflection component from the disk. 

This model, {\tt TBabs*(compps+bbodyrad)}, fitted the two 
spectra well, and yielded the best-fit parameters given in 
Table~\ref{tab:Xfit}.
Both spectra favored slightly smaller values of $R_\mathrm{bb}$ 
and $\tau$, if we assume $kT_\mathrm{e} = 100$ keV, 
compared with the case of $kT_\mathrm{e} = 20$ keV. 

We also tested the alternative possibility that the seed photons of 
Comptonization are provided by disk blackbody 
emission. Thus, the {\tt bbodyrad} component 
in the {\tt TBabs*(compps+bbodyrad)} model was 
replaced by {\tt diskbb}, and the inner disk temperature 
$kT_\mathrm{in}$ of {\tt diskbb} was set to be the same as 
$kT_\mathrm{seed}$ of {\tt compps} 
($kT_\mathrm{seed} = -kT_\mathrm{in}$, in XSPEC terminology). 
This {\tt TBabs*(compps+diskbb)} model was also found to fit the 
two spectra well, yielding comparable reduced $\chi^2$ values 
to those of {\tt TBabs*(compps+bbodyrad)}. 

Although we have assumed a spherical corona above, 
following previous works, 
we have confirmed that the choice of the coronal geometry 
does not affect the main conclusions from the 
{\tt TBabs*(compps+bbodyrad)} and {\tt TBabs*(compps+diskbb)} 
models. If a slab or cylindrical corona is assumed (i.e., {\tt geom} 
$=1$ or $2$, respectively), $\tau$ changes by a factor 
of $\lesssim$2, but the other free parameters remain unchanged 
within their 90\% error ranges. The chi-squared values 
were also found to depend little ($|\Delta \chi^2| \lesssim 1$) 
on the assumed coronal geometries.

We next investigated the spectral variation over a longer period. 
Figure~\ref{fig:XRTspecall} presents the other XRT 
spectra, in addition to those of March 27 and 31, which were 
already analyzed. The spectral profile varied in a complex 
manner: the peak energy of the soft component shifted 
toward higher energies from March 31 to April 6--7 and 
then moved back to lower energies in April 8--9, even though 
the X-ray flux was comparable among these three epochs. 
These XRT spectra were also well reproduced individually 
by the {\tt TBabs*(powerlaw+bbodyrad)} model.

Figure~\ref{fig:trend_fitpars} shows  
the best-fit parameters of these spectra 
in chronological order. It also 
shows the 3$\sigma$ upper limits of the 
unabsorbed 0.3--10 keV flux on April 4 and 5, 
calculated by assuming the best-fit model 
on April 6. As suggested by the spectral 
shape changes, the variations of 
$kT_\mathrm{bb}$ and $R_\mathrm{bb}$ 
are rather complex, and cannot be described as
a simple function of the X-ray flux. 

The spectra were also fit well with the {\tt TBabs*(powerlaw\\+diskbb)} model. The derived 
$kT_\mathrm{in}$ and $R_\mathrm{in}$ varied in 
a similar manner to $kT_\mathrm{bb}$ and 
$R_\mathrm{bb}$ of the {\tt TBabs*(powerlaw+bbodyrad)} 
model, respectively.

\section{Optical Observations and Multi-wavelength Spectral Energy Distributions} \label{SED}
\begin{figure}[htb]
\begin{center}
\includegraphics[width=7.5cm,bb=0 0 768 768]{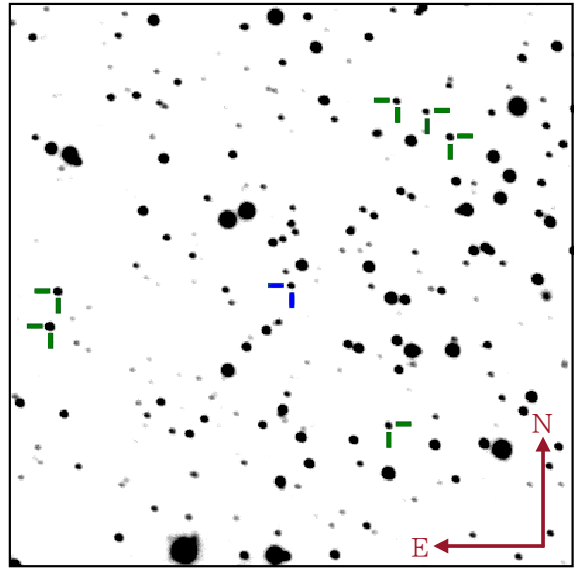}
\caption{An $R_C$-band finding chart for MAXI J1807$+$132, with 
a field of view of $4'.8 \times 4'.8$. The pair of blue bars at the center 
of the image point MAXI J1807$+$132, while green bars 
indicate the field reference stars.
\label{fig:imgopt}}
\end{center}
\end{figure}

Optical photometric observations of MAXI J1807$+$132 
were performed with the $g'$-, $R_C$-, and $I_C$-band 
filters for 4 nights from 2017 March 27 to 30, with the 
{\it Murikabushi} 105 cm telescope at the Ishigakijima 
Astronomical Observatory in Okinawa, Japan, and the 
{\it MITSuME} 50 cm telescope of Akeno Observatory in 
Yamanashi, Japan \citep[for detailed information, 
see][and reference therein]{tac17b}. 
The target was observed for $\sim$2~h on each day, 
during which simultaneous three-band observations 
were repeated, with the individual exposure times of 60 s. 
The raw data were preprocessed in a standard manner: 
subtraction of dark and bias, followed by flat fielding. 
The pixel coordinates were calibrated into celestial 
coordinates via WCSTools \citep{Min97}. After these 
treatments, we combined all the frames taken in a night 
in each band, and performed aperture photometry using 
IRAF tasks to estimate the magnitude of this object by 
comparing with six local reference stars.
Figure~\ref{fig:imgopt} shows the stacked $R_C$-band 
image obtained with the {\it Murikabushi} telescope on March 
27, where MAXI J1807$+$132 and the 6 reference stars 
are indicated. 

\begin{figure}[htb]
\epsscale{1.1}
\plotone{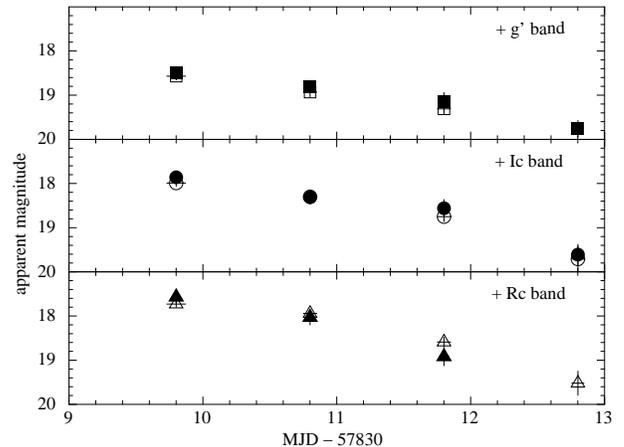}
\caption{The optical light curves of MAXI J1807$+$132 
in the $g'$, $R_C$, and $I_C$ bands. The open and filled 
symbols indicate the data from {\it Murikabushi} and 
{\it MITSuME} telescopes, respectively. \label{fig:LCopt}}
\end{figure}

The apparent magnitudes in the individual nights 
are plotted in Figure~\ref{fig:LCopt}, which clearly shows 
decay in all three bands, typically by $\sim$$+$0.4 
mag per day. Previously, the source was much fainter, 
at least by $\sim$3 mag, because multi-epoch 
Pan-STARRS observations gave an average 
$r$-band magnitude of $21.19 \pm 0.09$ mag \citep{den17}, 
which is $\sim$3 mag larger (i.e., the flux  
is $\sim$16 times lower) than the $R_C$-band 
magnitude estimated on March 27. 
This suggests that the emission from the companion 
star contributes only less than $\sim 6$\% of the total 
optical flux on March 27.

\begin{figure}[htb]
\epsscale{1.1}
\plotone{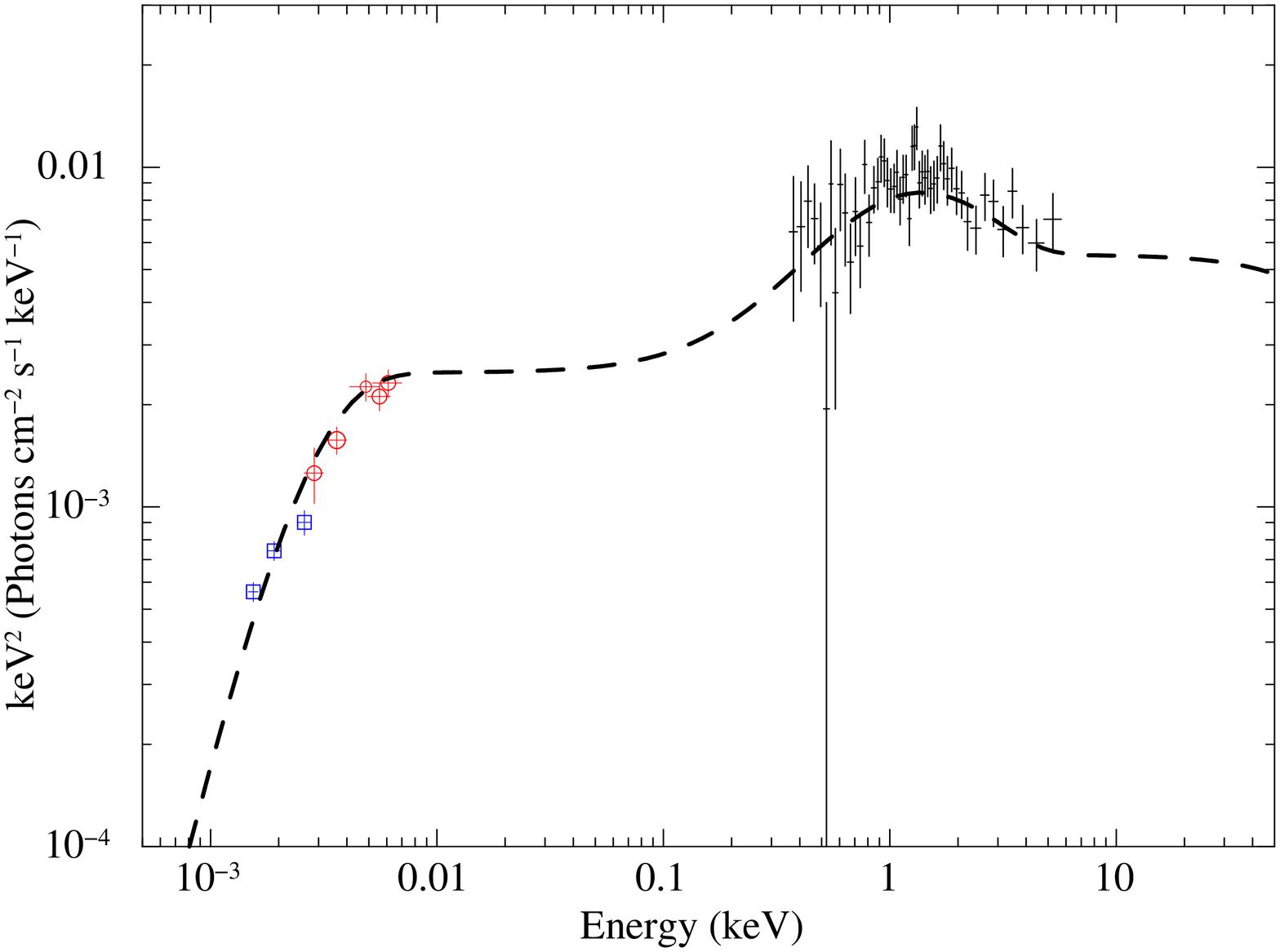}
\caption{A multi-wavelength SED of MAXI J1807$+$132 
on March 27 corrected for interstellar extinction. The 
{\it Swift}/XRT (black crosses), {\it Swift}/UVOT (red open 
circles), and the optical $g'$, $R_C$, and $I_C$-band data 
(blue open squares) from the {\it Murikabushi} telescope 
are plotted. The dashed line shows the best-fit {\tt diskir} model. 
\label{fig:SED}}
\end{figure}

Figure~\ref{fig:SED} shows the multi-wavelength 
spectral energy distribution (SED) on March 27,  
where the {\it Swift}/XRT and UVOT data are 
plotted together with those from the {\it Murikabushi} 
telescope. 
Here we examine the optical flux for a possibility 
of emission from the outer region of a standard accretion 
disk, irradiated by X-ray emission.
The illuminating source is ambiguous;
either a disk blackbody emission if 
the central object is a black hole, or a 
disk blackbody plus blackbody if a 
neutron star. However, there is no 
available model that includes blackbody 
emission. We thus account only for the 
disk blackbody emission here and 
attempt to fit the SED with an irradiated 
disk model ``{\tt diskir}'' \citep{gie08,gie09}.

The {\tt diskir} model calculates the 
emission from a standard disk, considering partial 
Comptonization of inner-disk photons. The outer region 
of the disk is assumed to be illuminated by a fraction 
of X-rays from the central region, to achieve a higher 
temperature and emissivity via reprocessing.
The free parameters of the model are the inner disk 
temperature $kT_\mathrm{in}$, the photon index $\Gamma$ 
and electron temperature $kT_\mathrm{e}$ of the 
Comptonized component, the luminosity ratio 
$L_\mathrm{C}/L_\mathrm{d}$ between the 
Comptonized corona and the disk, the fraction $f_\mathrm{in}$ 
of the luminosity of the Comptonized component that is 
thermalized in the inner disk, the fraction $f_\mathrm{out}$ 
of the bolometric flux that illuminates the outer disk, 
the radius $r_\mathrm{irr}$ of the Compton illuminated disk, 
the outer disk radius $R_\mathrm{out}$, and the normalization, 
depending on the inner disk radius $R_\mathrm{in}$ and 
the distance in the same manner as {\tt diskbb}.  
Following previous works \citep{gie08, gie09}, we set 
$r_\mathrm{irr} = 1.1 R_\mathrm{in}$\footnote{We have 
confirmed that the choice of the $R_\mathrm{irr}$ value does 
not strongly affect the estimation of $f_\mathrm{out}$. 
The resultant $f_\mathrm{out}$ value was kept 
unchanged within its 90\% confidence range, in the 
case of $R_\mathrm{irr}=1.1$, 5, and 10.}. 
Considering the results of our XRT 
spectral analysis, we assumed $\Gamma = 2.0$, 
$kT_\mathrm{e}=100$ keV, $f_\mathrm{in} = 0.1$, 
and left the other parameters free to vary. 
To account for the optical extinction, the 
{\tt redden} model with $E(B-V) = 0.13$ (which is converted 
to $N_\mathrm{H} \sim 1 \times 10^{21}$ cm$^{-2}$ via 
the relation given in \citealt{pre95}) was multiplied to 
{\tt diskir}. 

As shown in Figure~\ref{fig:SED}, the overall 
SED profile has been reasonably well reproduced by the 
{\tt diskir} model with $kT_\mathrm{in} \approx 0.4$ keV, 
$L_\mathrm{C}/L_\mathrm{d} \approx 2.3$, 
$R_\mathrm{in} \approx 3$ km, $R_\mathrm{out} 
\approx 1 \times 10^5 R_\mathrm{in} \approx 3 \times 10^5$ 
km (where the distance and inclination are assumed as 5 kpc 
and $0^\circ$, respectively),
and $f_\mathrm{out} \approx 3.7 \times 10^{-2}$. 
The estimated values of $kT_\mathrm{in}$ and 
$R_\mathrm{in}$ are comparable with those obtained from 
the XRT data alone using the {\tt diskbb+powerlaw} 
model (Section~\ref{xrtana}).

\section{Discussion} \label{discussion}
\subsection{The Nature of MAXI J807$+$132}
We studied the behavior of the new X-ray transient 
MAXI J1807$+$132 using the multi-wavelength data 
of the MAXI/GSC, {\it Swift}, and optical telescopes. 
The source is likely to be identified with  
2MAXIt J1807$+$132, which is listed in the first MAXI/GSC 
transient source catalog \citep{kaw16} based on  
a long-term X-ray brightening event in 2011 May. 
Although \citet{kaw16} primarily aimed at a search 
for tidal disruption events (TDEs) by extragalactic 
supermassive black holes, the 2011 May episode 
of 2MAXIt J1807$+$132 was not categorized therein 
as a TDE. Below, we consider various interpretations 
of the nature of this object.

\subsubsection{A Tidal Disruption Event?}
\citet{kaw16} concluded that 2MAXIt J1807$+$132 
is unlikely a TDE, because it has exhibited 
multiple enhancements (though with lower 
significances, at $\sim$2$\sigma$ levels) from 
2009 to 2013.
Identifying the present source MAXI J1807$+$132 
with 2MAXIt J1807$+$132, the TDE interpretation 
becomes even less likely, because 
the interval of the two strongest flaring events is much 
shorter than those predicted for TDEs \citep[typically 
$\sim$$10^4$ to $\sim$$10^5$ years; e.g.,][]{kaw16}. 

\begin{figure}[htb]
\epsscale{1.1}
\plotone{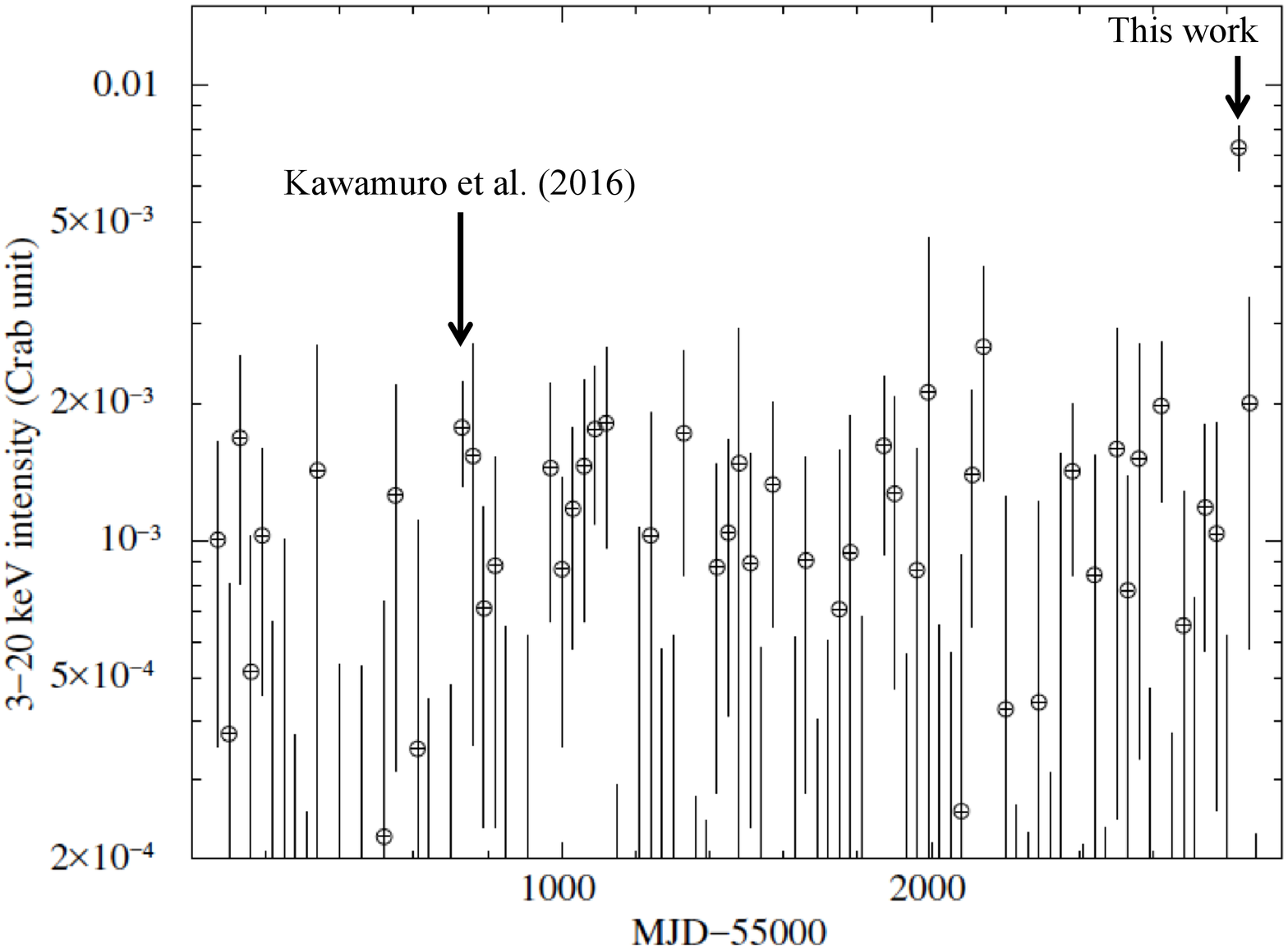}
\caption{A 2--20 keV MAXI/GSC light curve of MAXI J1807$+$132 
for the last $\approx$7.5 years, with 30-day time bins. 
The thick solid arrows 
indicate the flares reported by \citet{kaw16} and this work. 
\label{fig:maxilc_all}}
\end{figure}

To search for other brightening episodes of MAXI J1807$+$132 
($=$2MAXIt J1807$+$132), we analyzed the entire 
MAXI/GSC data of this sky region using the on-demand 
process system\footnote{\url{http://maxi.riken.jp/mxondem/}}. 
Figure~\ref{fig:maxilc_all} shows the obtained light 
curve, from 2009 August to 2017 April. 
However, we detected no X-ray enhancements with 
a significance of $>$3$\sigma$, other than the flares 
in 2011 and 2017. As noticed in Fig.~\ref{fig:maxilc_all}, 
the present flaring event in 2017 is the brightest one in 
the last 7.5 years.

The {\it Swift}/XRT data allow us to further argue against 
the TDE interpretation of the present object.
The XRT spectra of MAXI J1807$+$132 in 2017 March 
and April have been well described with a soft thermal 
component (blackbody or disk blackbody) visible 
below $\sim$2 keV,  and a hard tail with a photon 
index of $\sim$2. The thermal component has a temperature 
higher than those of TDEs \citep[typically 
$\lesssim$0.1 keV;][]{esq08,mak10}. The hard 
X-ray tail is much stronger than those of non-jetted 
TDEs, although it could be compatible with the 
spectrum of the jetted TDE Swift J164449.3$+$573451 
\citep{bur11, blo11, lev11}.

Typical TDEs, either with or without jets, are considered to 
decay with time $t$ as $\propto  t^{-5/3}$ \citep[e.g.,][]{ree88, phi89}.
In contrast, the X-ray emission of MAXI J1807$+$132 
in the present flaring event decayed much more rapidly;
fitting the XRT light curve (Fig.~\ref{fig:trend_fitpars}a) 
with a power-law model, 
we obtained the best-fit decay curve as $\propto  t^{-6 \pm 1}$, 
where $t$ is the time since 2017 March 13 
when the source was first recognized with MAXI. 
This fast decay also shows that the source is unlikely a TDE.

\subsubsection{A Galactic Magnetic Cataclysmic Variable (CV)?}
Having excluded the TDE interpretation, we hereafter assume 
that the source is a binary system located in our galaxy. 
One of the most abundant subclasses of such binaries 
is accreting magnetic CVs (Polars and Intermediate Polars).
Indeed, the X-ray flux observed on March 31, 
$\sim 7 \times 10^{-12}$ erg cm$^{-2}$ s$^{-1}$, translates 
to a source luminosity of $\sim 10^{34}$ erg s$^{-1}$,
if the source distance is assumed, e.g., to be 5 kpc.
This luminosity is typical of magnetic CVs during their 
brightening episodes \citep[e.g.,][]{rev08}.
An X-ray spectrum of a typical CV is composed of 
an optically thin thermal plasma emission produced 
in the accretion columns, and a blackbody emission 
from the polar cap region of the white dwarf.
We found that the XRT spectra of MAXI J1807$+$132
above 1 keV can be fit with the optically thin plasma
model {\tt mekal} with a temperature of $\gtrsim$ 5 keV,
which is consistent with those of typical magnetic CVs. 
However, the soft thermal component of MAXI J1807$+$132
has a much higher temperature than the blackbody 
component in CVs ($<$0.1 keV).
Therefore, the magnetic CV interpretation is unlikely.

\subsubsection{A High Mass X-ray Binary (HMXB)?}
Since MAXI J1807$+$132 is located at a relatively 
high Galactic latitude of $15^\circ$, the possibility of its 
being a HMXB would be low. If the companion star was  
an O or B-type star located at $\sim$8 kpc, 
with an absolute $V$-band magnitude of $-4$ to $0$ 
mag, we should have easily detected it with an apparent  
magnitude of 10--15 mag, even when 
the system was not active. In addition, during the 
present X-ray brightening, the optical flux of MAXI J1807$+$132 
increased by an order of magnitude above the quiescence level.
Such optical brightening should not take place in HMXBs
even during increased mass accretion rates, because 
their optical fluxes are dominated by those from the 
mass-donating companions rather than from outer 
accretion disks. Finally, MAXI J1807$+$132 exhibited 
spectra that are considerably softer than those of high 
mass X-ray pulsars, which are roughly described by a 
power-law with a photon index of $\lesssim$1 \citep{cob02}.
Therefore, MAXI J1807$+$132 cannot be a HMXB,
regardless of the nature of the compact object involved.

\subsubsection{A transient BHXB?}
The long-term spectral variation of MAXI J1807$+$132 
throughout the present outburst is qualitatively similar to 
those seen in transient BHXBs. The spectrum became 
softer from March 26--29 to March 31 and harder again 
in April. Under the limited statistics of the XRT data,
this behavior could be explained if we were witnessing 
a hard-to-soft and soft-to-hard transitions in these periods, 
respectively. Therefore, below we examine in more details
whether this interpretation is feasible or not.

If the source is a BHXB as assumed above, the prominent, 
low-temperature component seen on March 31 should be 
explained as disk emission in the soft state or an intermediate 
state (if in the hard state, the disk emission would not be 
as bright as in the March 31 spectrum). 
Indeed, similar temperature and strength of the soft 
component have been obtained in other BHXBs 
\citep[e.g.,][]{nak14}. However, if this were the case, 
MAXI J1807$+$132 should have an unusually large 
distance. Taking into account that the soft-to-hard 
transition of BHXBs normally occurs at a few \% Eddington 
luminosity \citep{mac03}, which corresponds to $\sim 10^{37}~
(M_\mathrm{BH}/10~M_\sun)$ erg s$^{-1}$ ($M_\mathrm{BH}$ 
being the black hole mass), the March 31 flux as quoted above 
($\sim$7 $\times 10^{-12}$ erg cm$^{-2}$ s$^{-1}$) implies  
a source distance as large as $\sim$100 
$(M_\mathrm{BH}/10~M_\sun)^{1/2}$ kpc. 

There is yet another evidence against the BHXB 
interpretation. Applying the {\tt diskbb+powerlaw} model to the 
March 27 spectrum, we obtained an unusually small inner disk 
radius, $R_\mathrm{in} \, \sim 1 (\frac{\cos i}{\cos 0^\circ})^{-1/2} 
(D/5$ kpc$)$ km. In order to identify it with the radius of the 
innermost circular orbit, we would have to assume an extreme 
inclination, e.g., $i \gtrsim 85^\circ$, and a very light black 
hole (e.g., $\sim$3 $M_\sun$) with a substantial spin.

Modeling the multi-wavelength SED on March 27 
with the {\tt diskir} model (Section~\ref{SED}) yielded 
an irradiation fraction of $f_\mathrm{out} \approx 3.7 \times 
10^{-2}$. This value is about several to ten times larger 
than those obtained from black hole X-ray binaries with 
X-ray luminosities of $\gtrsim 10^{35}$ erg s$^{-1}$
(e.g., \citealt{gie08, gie09, chi10, shi13, nak14}, but see 
\citealt{rah12}). Even if no extinction is assumed (where 
the optical flux changes only by a factor of $\approx$2), 
the value of $f_\mathrm{out}$ is reduced only by a factor 
of $\lesssim$3, and is still somewhat larger than that of 
black hole X-ray binaries. The situation worsens if we 
assume the stronger reddening, $E(B-V) = 0.28$, 
estimated via optical spectroscopy \citep{mun17}. 
Considering all these results, we conclude 
that the source is unlikely to be a BHXB.

\subsubsection{A Neutron Star LMXB?} \label{sec:lmxbint}
A remaining possibility is that the source is 
a neutron star LMXB. The XRT spectra resemble those 
of neutron star LMXBs in their dim phases, with a luminosity 
below $\sim 10^{35}$ erg s$^{-1}$; 
at this luminosity range, they often exhibit a 
prominent soft thermal component with a power-law tail 
\citep{asa96, wij01, jon05, sak14, cha14}, 
whereas such an apparent two-component feature 
is less significant when they are in the more luminous 
hard state \citep[e.g.,][]{bar01, lin07, arm17b}. 
In particular, the properties of the XRT spectrum on 
2017 March 27 are similar to those obtained in the two 
{\it Suzaku} observations of Aql X-1 in its dim phases 
(``Obs 5'' and ``Obs 6'' in \citealt{sak14}), as noticed 
by comparing the {\tt compps} results. 
If MAXI J1807$+$132 is a neutron star LMXB, 
and if the unabsorbed 0.8--100 keV luminosity on 
March 27 was between those in ``Obs 5'' and ``Obs 6'' 
of Aql X-1 ($5 \times 10^{35}$ erg s$^{-1}$ and 
$1 \times 10^{34}$ erg s$^{-1}$, respectively), 
the distance is calculated as  $D \sim$1--8 kpc. 

A similar distance, $D=5$ kpc, is derived from the relation 
between the luminosity versus the photon index 
for neutron star LMXBs \citep{wij15}, by using 
the photon index on March 27 ($\Gamma \approx 2.4$, 
when a single power-law model is applied) and the 
unabsorbed flux, $4 \times 10^{-11}$ erg cm$^2$ sec$^{-1}$. 
Assuming $D = 5$ kpc, the 3$\sigma$ upper limit of 
the unabsorbed 0.3--10 keV flux on April 5, 
$8.8 \times 10^{-13}$ erg cm$^{-2}$ s$^{-1}$, 
which is the lowest flux constraint in the XRT datasets, 
is converted to the Eddington ratio of 
$\approx$$1.5 \times 10^{-5}$, 
for a neutron star with a mass of $1.4 M_\sun$.
This value is consistent with the minimum luminosity 
for neutron star LMXBs determined by \citet{tom05}. 

Such a low luminosity (below $10^{35}$ erg s$^{-1}$) would be 
favored to explain the high $L_\mathrm{OPT}/L_\mathrm{X}$ ratio, 
where $L_\mathrm{OPT}$ and $L_\mathrm{X}$ are the optical 
and X-ray luminocities, respectively, given a correlation of 
$L_\mathrm{OPT} \propto L_\mathrm{X}^{\alpha}$ with 
$\alpha \sim 0.5$ \citep{van94, rus06, rus07}. The X-ray 
flux of MAXI J1807$+$132 decreased by $\sim$2 orders of 
magnitude from March 26 to the early April. Similar rapid 
flux decay has been observed in other neutron star LMXBs, 
such as Aql X-1, 4U 1608$-$52, and MAXI J1421$-$613 
\citep{cam98,asa13,ser15}, at luminosities below $10^{36}$ erg s$^{-1}$, 
where the propeller effect is considered to start operating,   
and the centrifugal force prevents steady accretion onto 
the neutron star \citep{mat13}. 
These results provide yet another support to our 
identification 
of MAXI J1807$+$132 with a neutron star LMXB in 
a dim phase.

\subsection{Physical Interpretation of the X-ray Spectra and Their Variations}
Let us examine the X-ray properties of  MAXI J1807$+$132,
assuming that it is a dim neutron star LMXB. The soft X-ray 
component of such an object,
at luminosities below $\sim 10^{35}$ erg s$^{-1}$, is generally 
considered as thermal emission from the surface of the neutron 
star. 
The small radius of the blackbody emission region
(a few km) can be naturally explained if only a part of
the surface radiates X-rays. In fact, the blackbody 
radius of Aql X-1 was found to decrease from 
$\sim$10 km at $\sim$$1 \times 10^{36}$ erg s$^{-1}$
down to $\sim$3 km at  $\sim$$1 \times 10^{34}$ 
erg s$^{-1}$ (Figure 6 of \citealt{sak14}), presumably 
because of the appearance of weak magnetic fields
which limit the accretion flows to the magnetic poles.
By contrast, the origin of the hard power-law tail is not yet 
fully understood. It allows several different interpretations, 
such as Comptonization of the blackbody emission in a hot 
accretion flow \citep{sak14}, bremsstrahlung from 
the hot flow itself \citep{cha14}, 
and the jet emission \citep{fen03}. In Section~\ref{xrtana}, we 
applied the {\tt compps} model to the XRT spectra on March 
27 and 31, to examine the first interpretation in comparison 
with the {\it Suzaku} results of Aql X-1 \citep{sak14}, and 
found similarities in their best-fit parameters 
(see also Section~\ref{sec:lmxbint}). 

The contribution by bremsstrahlung from the Comptonized 
corona was evaluated quantitatively by \citet{ono17}, based on the 
observationally estimated accretion flow geometry in Aql X-1. 
When the source luminosity is $>5 \times 10^{36}$ erg s$^{-1}$,
they estimated that the bremsstrahlung luminosity is 
$L_\mathrm{Br} \sim 1 \times 10^{34}$ erg s$^{-1}$. Then, 
assuming that MAXI J1807$+$132 has a typical luminosity 
of $2 \times 10^{35}$ erg s$^{-1}$, and that $L_\mathrm{Br}$ 
decreases as the mass accretion rate gets lower, we 
conclude that $L_\mathrm{Br}$ is likely to be still lower
than the luminosity of the power-law tail.

We have detected significant variations in the temperature 
and the radius of the emission region of the blackbody 
component, which are not determined by the X-ray 
luminosity alone (see Fig.~\ref{fig:XRTspecall} 
and Fig.~\ref{fig:trend_fitpars}). Similar peculiar behavior 
has been observed in other LMXBs, like Aql X-1 
\citep{rut02} and XTE J1709$-$267 \citep{jon05}, 
at luminosities of $10^{33}$--$10^{35}$ erg s$^{-1}$.  
These variations were suggested to arise from residual 
accretion onto the polar-cap region of the neutron star, 
in association with neutron star cooling 
\citep[e.g.,][]{cac10, deg12}. In the case of 
MAXI J1807$+$132, however, it is unclear 
whether the neutron star was at the stage 
of crustal cooling during the {\it Swift} observations, 
provided that the timescale of its flux decay was somewhat 
shorter than those in other neutron star LMXBs in quiescence 
\citep{hom15}. 

The spectra of MAXI J1807$+$132 could be 
categorized into two types: (1) a flatter profile with a higher 
$kT_\mathrm{bb}$ and a smaller $R_\mathrm{bb}$ (e.g., 
the March 27 spectrum), and (2) a more complex profile 
in which the soft component, with a lower $kT_\mathrm{bb}$ and 
a larger $R_\mathrm{bb}$, and the hard tail are distinctive 
(e.g., the March 31 spectrum). This bimodality cannot be 
attributed to luminosity changes, since the spectrum 
switched a few times between the two types as the overall X-ray 
luminosity decayed almost monotonically.

\begin{figure}[htb]
\epsscale{1.1}
\plotone{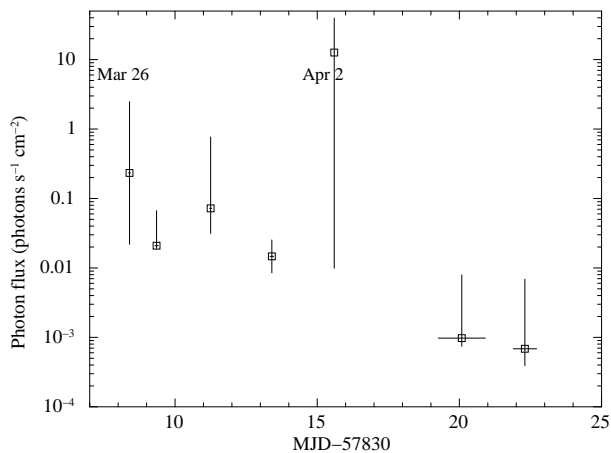}
\caption{Variation of the 0.01--100 keV photon flux 
of the soft thermal emission, calculated 
with the {\tt tbabs*(bbodyrad+compps)} model.
The abscissa is the same as Fig.~\ref{fig:trend_fitpars}. 
\label{fig:trend_pflux}}
\end{figure}

One possible explanation of the observed variations 
between the two types would be to assume that 
the seed blackbody spectrum suffers strong color 
hardening, by a factor of $\kappa \sim 3$, 
in type-(1) spectra, whereas such an effect is 
small in type-(2) spectra so that the seed spectrum 
is close to a ``bare'' blackbody. In fact, \citet{tak09}
observed spontaneous fluctuations in $\kappa$ 
in the LMXB 4U 1608$-$52 (even though the 
effect was only $\sim$20\% and was observed 
in the disk emission of the soft state). 
To examine this interpretation, we investigated 
the trend in the total photon flux of the soft thermal 
component, including both the direct and 
Compton-scattered components, assuming  
isotropic emission and conservation of the 
number of photons in Comptonization.  
In this analysis, the {\tt TBabs*(bbodyrad+compps)} 
model was applied to the individual XRT spectra 
with $kT_\mathrm{e}=$ 100 keV. 
As shown in Figure~\ref{fig:trend_pflux}, 
the photon flux decreased rather monotonically. 
Thus, the observed variations could be described 
by a change in the color hardening factor 
(for some unspecified reasons) during a 
monotonic decrease in the mass accretion rate.

\subsection{X-ray and Optical Flux Correlation}
The optical flux was reduced from March 27 
to 31 by $\sim$0.4 mag per day as the 
X-ray flux decreased. The optical decay 
can be expressed as $F_\mathrm{OPT} 
\propto \exp(-t/\beta_\mathrm{OPT})$, where 
$F_\mathrm{OPT} $ represents the optical 
flux and $\beta_\mathrm{OPT} \sim 2.7$ day. Fitting the XRT 
0.3--10 keV light curve (Fig.~\ref{fig:trend_fitpars}) 
in March 26--31 with an exponential function, 
$F_\mathrm{X} \propto \exp(-t/\beta_\mathrm{X})$, 
we obtain $\beta_\mathrm{X} = 2.4 \pm 0.5$. 
From these two functions, the X-ray versus 
optical flux relation is derived as 
$F_\mathrm{OPT} \propto F_\mathrm{X}^{\alpha}$, 
where $\alpha \sim$0.7--1.1. This $\alpha$ value 
is comparable with those obtained for other neutron 
star LMXBs in their dim ($L_\mathrm{X} \lesssim 10^{36}$ 
erg s$^{-1}$) periods \citep{van94, rus06, rus07}, which 
have been explained by reprocessing of X-ray irradiation 
in the outer accretion disk. We also find that the optical 
and X-ray luminosities on March 27, estimated by 
assuming a distance of 5 kpc, lie on the trend in 
$L_\mathrm{X}$ and $L_\mathrm{OPT}$ of 
neutron star LMXBs in \citet{rus07}. 
These results further reinforce the LMXB interpretation 
of MAXI J1807$+$132.

\acknowledgments
We are grateful to the anonymous referee for constructive 
comments. MS acknowledges support by the Special Postdoctoral 
Researchers Program at RIKEN. This work is partly supported 
by a Grant-in-Aid for Young Scientists (B) 16K17672 (MS), for 
JSPS Fellows for young researchers (YT, TK), and for Scientific 
Research 17K05384 (YU) and 16K05301 (HN). 
This research has made use of MAXI data provided by RIKEN, 
JAXA and the MAXI team and {\it Swift} data supplied by the UK Swift 
Science Data Centre at the University of Leicester. This 
work was partially carried out by the joint research 
program of the Institute for Cosmic Ray Research (ICRR), 
University of Tokyo.

\vspace{5mm} 
\facilities{MAXI (GSC), Swift (XRT, UVOT)}
\software{UK Swift Science Data Centre tools \citep{eva09}, HEASOFT, XSPEC \citep{arn96}}

\end{document}